\documentclass[journal,transmag]{IEEEtran}
\ifCLASSINFOpdf
\else
\fi

\usepackage{amsthm}
\usepackage[cmex10]{amsmath}
\usepackage{amsfonts}
\usepackage{amssymb}

\ifCLASSINFOpdf
\usepackage[pdftex]{graphicx}
\usepackage{subfigure}
\else

\usepackage{graphicx}
\usepackage{subfigure}
\fi
\usepackage{epstopdf}
\usepackage{color,xcolor}
\usepackage{multicol}  
\usepackage{multirow}
\usepackage{booktabs} 
\usepackage{threeparttable}
\usepackage{algorithmic,algorithm}
\usepackage{subfigure}

\usepackage[font=normalsize]{caption}

\usepackage{algorithmic,algorithm}

\usepackage[numbers,sort&compress]{natbib}
\usepackage{color}

\usepackage{url}

\usepackage{mdwtab}

\begin{document}
%
\title{PacketCGAN: Exploratory Study of Class Imbalance for Encrypted Traffic Classification Using CGAN\\ IEEE \textsc{Transactions on Network and Service Management}}


\author{\IEEEauthorblockN{Pan Wang\IEEEauthorrefmark{1},~\IEEEmembership{Member,~IEEE}, 
Shuhang Li\IEEEauthorrefmark{1},
Feng Ye\IEEEauthorrefmark{2},~\IEEEmembership{Member,~IEEE}, 
Zixuan Wang\IEEEauthorrefmark{1}, and
Moxuan Zhang\IEEEauthorrefmark{3}}
\IEEEauthorblockA{\IEEEauthorrefmark{1}School of Modern Posts, Nanjing University of Posts \& Telecommunications, Nanjing, China}
\IEEEauthorblockA{\IEEEauthorrefmark{2}Department of Electrical \& Computer Engineering, University of Dayton, Dayton, OH, USA}
\IEEEauthorblockA{\IEEEauthorrefmark{3}Schools of International Education, Jinling Institute of Technology, Nanjing, China}
\thanks{Manuscript received December 1, 2019. 
Corresponding author: Pan Wang (email: wangpan@njupt.edu.cn).}}

\markboth{Journal of \LaTeX\ Class Files,~Vol.~14, No.~8, August~2015}%
{Shell \MakeLowercase{\textit{et al.}}: Bare Demo of IEEEtran.cls for IEEE Transactions on Magnetics Journals}
%



\IEEEtitleabstractindextext{%
\begin{abstract}
With more and more adoption of Deep Learning (DL) in the field of image processing, computer vision and NLP, researchers have begun to apply DL to tackle with  encrypted traffic classification problems. Although these methods can automatically extract traffic features to overcome the difficulty of traditional classification methods like DPI in terms of feature engineering, a large amount of data is needed to learn the characteristics of various types of traffic. Therefore, the performance of classification model always significantly depends on the quality of datasets. Nevertheless, the building of datasets is a time-consuming and costly task, especially encrypted traffic data. Apparently, it is often more difficult to collect a large amount of traffic samples of those unpopular encrypted applications than well-known, which leads to the problem of class imbalance between major and minor encrypted applications in datasets. In this paper, we proposed a novel traffic data augmenting method called PacketCGAN using Conditional GAN. As a generative model, PacketCGAN exploit the benefit of CGAN to generate specified traffic to address the problem of the datasets' imbalance. As a proof of concept, three classical DL models like Convolutional Neural Network (CNN) were adopted and designed to classify four encrypted traffic datasets augmented by Random Over Sampling (ROS), SMOTE(Synthetic Minority Over-sampling Techinique) , vanilla GAN and PacketCGAN respectively based on two public datasets: ISCX2012 and USTC-TFC2016. The experimental evaluation results demonstrate that DL based encrypted traffic classifier over dataset augmented by PacketCGAN can achieve better performance than the others.

\end{abstract}

\begin{IEEEkeywords}
encrypted traffic classification, data augmentation, Conditional Generative Adversarial Network, traffic identification, class imbalance.
\end{IEEEkeywords}}

\maketitle

\IEEEdisplaynontitleabstractindextext

%
\IEEEpeerreviewmaketitle

\section{Introduction}
%
%
%
%
\IEEEPARstart{W}{ith} the rapid development of network technology, the types and quantity of traffic data in cyberspace are increasing. Network traffic identification and classification is a crucial research task in the area of network management and security. It is the footstone of dynamic access control, network resources scheduling, content based billing, intrusion and malware detection etc. High efficient and accurate traffic classification is of great practical significance to provide service quality assurance, dynamic access control and abnormal network behaviors detection. With the widespread adoption of encryption techniques for internet, especially 5G and IoT applications, the growth of portion of encrypted traffic has dramatically posed a huge challenge for QoS, network management and security monitoring. Therefore, studies on encrypted traffic classification not only help to improve the fine-grained network resource allocation based on application, but also enhance security level of network and applications.

Traditionally, the evolution of encrypted traffic classification technology has gone through three stages: port matching, payload matching and flow statistical characteristics based classification methods. Port matching based method infers applications' types by assuming that most applications consistently use 'well known' TCP or UDP port numbers, however, the emergence of port camouflage, dynamic port, proprietary protocols with user-defined ports and tunneling technology makes these methods lose efficacy quickly. Payload matching based methods, namely, DPI (Deep Packet Inspection) technology cannot deal with encrypted traffic because of invisible packet content of encrypted traffic, in addition, it incurs high computational overhead and requires manual signatures maintenance~\cite{Finsterbusch2014,SDN-HGU,PhoneNumber}. As a result, in order to attempt to solve the aboved problems of encrypted traffic identification , flow-based methods emerged, which usually combine statistical or time series traffic features with Machine Learning (ML) algorithms,  such as naive bayes(NB), support vector machine(SVM), decision tree, Random Forest(RF), k-nearest neighbor(KNN) ~\cite{Pescape2008,Sun2010,Velan2015,Arndt2011}. Although classical machine learning approaches can solve many issues that port and payload based methods cannot solve, it still has some limitations, such as handcrafted traffic features driven by domain-expert, time-consuming, lack of ability of automation, rapidly outdated when compared to the evolution. Unlike most traditional ML algorithms, DL performs automatic feature extraction without human intervention, which undoubtedly makes it a highly desirable approach for traffic classification, especially encrypted traffic.  Recent research work has demonstrated the superiority of DL methods in traffic classification~\cite{MobileTC-2018}, such as MLP~\cite{Datanet}, CNN~\cite{deeppacket,Wang-1D-CNN,Wang2D-CNN,Seq2Img,HierarchicalTC}, SAE~\cite{blackhat}, LSTM~\cite{IoT-CNN-2017,HAST-IDS}.

However, due to the different popularity of various applications, the class imbalance problem of traffic samples often occurs when building traffic datasets. That is, the number of popular applications samples is much larger than others, which always leads to the misclassifying problems of minor applications and thereby incurs deterioration of classifier performance. Imbalanced class distribution of a dataset has posed a serious challenge to most ML based classifiers which assume a relatively balanced distribution~\cite{Japkowicz:2002}. Network traffic classification is no exception due to the imbalanced property of network traffic data~\cite{Vu2017,Vu2016}, especially encrypted traffic. Therefore, it is very crucial to address such challenges of imbalanced class distribution of traffic datasets for network traffic classification. However, there are very few studies focusing on traffic data augmentation for traffic classification to overcome the limitation of class imbalance.

In this paper, we proposed traffic data augmentation method called PacketCGAN using Conditional GAN, one of a genre of GAN, which can control the modes of the data to be generated. As a generative model, PacketCGAN exploits the benefit of CGAN to generate synthesized traffic samples by learning the characteristics of the original traffic data. The synthesized data is then combined with the original (viz. real) data to build the new traffic dataset and thereby keep balance between major and minor classes of the dataset. As a proof of concept, three classical DL models like CNN were adopted and designed to classify four types of encrypted traffic datasets augmented by ROS, SMOTE, vanilla GAN and PacketCGAN respectively using two public datasets: ISCX2012 and USTC-TFC2016. The experimental evaluation results demonstrate that DL based encrypted traffic classifier over our new dataset augmented by  PacketCGAN can achieve better performance than the other three in terms of encrypted traffic classification.

The rest of this paper is organized as follows. Section II introduces related works of traffic classification and some current methods for tackling with the problem of imbalanced class data. Section III describes the principles of GAN and CGAN. Section IV illustrates the methodology of PacketCGAN. The experimental results are provided and discussed in Section V. Section VI  concludes our work and presents some future works.

\section{Related works}\label{sec:related_works}
\subsection{ML and DL based approach of Traffic Classification}
Different from port and payload matching methods, ML based classification methods always use payload-independant parameters such as packet length, inter-arrival time and flow duration to circumvent the problems of encrypted content and user's privacy~\cite{2007legal}. Many work was carried out using ML algorithms during the last decades. In general, there are two learning strategies used: one is the supervised methods like decision tree, SVM and Naive Bayes, the other is unsupervised approaches like k-means and PCA~\cite{2008NguyenTC}. Nevertheless,  many drawbacks hindered ML based methods widely applied to traffic classification, such as handcrafted traffic features driven by domain-expert, time-consuming, unsuited to automation, rapidly outdated when compared to the evolution. Unlike most traditional ML algorithms, Deep Learning performs automatic feature extraction without human intervention, which undoubtedly makes it a highly desirable approach for traffic classification, especially encrypted traffic.  Recent research work has demonstrated the superiority of DL methods in traffic classification~\cite{MobileTC-2018,deeppacket,Datanet,Wang-1D-CNN,Wang2D-CNN,IoT-CNN-2017,HAST-IDS,Seq2Img,blackhat,TC-VAE,HierarchicalTC}. The workflow of DL based classfication usually consists of three steps. First, model inputs are defined and designed according to some principles, such as raw packets, PCAP files or flow statistics features. Second, models and algorithms are elaborately chosen according to models' characteristics and aim of the classifier. Finally, the DL classifier is trained to automatically extract the features of traffic.

\subsection{Traditional methods for handling imbalanced data}\label{subsec_imbalance}
In general, there are three methods for dealing with class imbalance problem: \emph{\textbf{Modifying the objective cost function, Sampling and Generating artificial data}}~\cite{Vu2016}. The approach of \emph{\textbf{modifying objective cost function}} alleviates the problem by means of weighting differently the data samples in minor and major classes, which gives higher score on the minor samples to penalize more intensely on miss-classifiying of the sample in the minor class. Sampling methods include two different ways of \emph{\textbf{under-sampling}} and \emph{\textbf{over-sampling}}, which is to reduce the size of major class by removing some major data samples and raise the ones in the minor class, respectively. Random under sampling (RUS) and Random over sampling (ROS) are two main methods of under-sampling and over-sampling~\cite{ROS2000}. RUS randomly removes some instances in major class, accordingly, ROS generates some copies of samples of the minor class. However, overfitting problem is always the main drawback of ROS due to generating same copies from the minor class. A classical method for generating artificial data is Synthetic Minority Over-sampling Techinique (SMOTE) in which minority samples are generated by synthetic samples rather than copies~\cite{SMOTE2002}.

\subsection{The application of GAN and other DL techniques in generating traffic data samples}
Due to the great success of GAN applying in images, computer vision and NLP etc., this innovative technique has been already applied to network security recently. A few current studies have shown that GAN has been applied in IDS and Malware detection to generate adversarial attacks to deceive and evade the detection systems~\cite{Lin2018IDSGANGA,Hu2017GAN} and thus effectively improve the performance of malware detection or IDS~\cite{Hu2017GAN,KimGAN,Lin2018IDSGANGA,Salem2018GAN,Knife-Fight}. Correspondingly, as for traffic classification, some researchers have introduced some approaches based on GAN to generate the traffic samples to overcome the imbalaced limitation of network data. In~\cite{Vu-GAN}, the authors proposed a novel method called auxiliary classifier GAN (AC-GAN) to generate synthesized traffic samples for balancing between the minor and major classes over a well-known traffic dataset NIMS. The AC-GAN took both a random noise and a class label as input in order to generate the samples of the input class label accordingly. The experimental results has shown that their proposed method achieved better performance compared to other methods like SMOTE. However, the NIMS dataset was only composed of  SSH and non-SSH two classes. In ~\cite{Hasibi2019AugmentationSF}, the authors proposed a novel data augmentation approach based on the use of Long Short Term Memory (LSTM) network to learn the traffic flow patterns and Kernel Density Estimation (KDE) for replicating the sequence of packets in a flow for classes with less population. The results have shown that this method can improve the performance of DL algorithms over augmented datasets. 

\section{Generative Adversarial Networks}\label{gan}
\subsection{GAN}\label{AA}
As an unsupervised learning model, a classic GAN network consists of two parts, the generator $G$ and the discriminator $D$. 
The role of the generator is to take random noise as input by learning the characteristic distribution of real data. The discriminator aims at determining whether the data is real or generated by $G$.  
The generator $G$ simulates the feature distribution $P_g$ of the real data by the prior distribution $P_z(z)$. The input of the discriminator is the real and generated data, correspondingly, the output $D(x)$ indicates the probability of whether the input data is real or not~\cite{r11}. During the training process, $G$ and $D$ play a two-player mini-max game until $D$ can't judge whether the sample data is real, which means that the two networks reach the Nash Equilibrium. The objective function of GAN can be expressed by \eqref{GAN}:
\begin{equation}
	\begin{split}
		\label{GAN}
		\mathop {\min }\limits_{\rm{G}} \mathop {\max }\limits_D V(D,G) = \mathbb{E}_{x\sim{p_{data}}(x)}[\log D(x)] \\
		+ {\mathbb{E}_{z\sim{p_z}(z)}}[\log (1 - D(G(z)))]
	\end{split}
\end{equation}

In Equation \eqref{GAN}, $P_{data}(x)$ represents distribution of the real data. When training $D$, the goal is to optimize the probability of TRUE $D(G(z))$ as small as possible and the probability of TRUE $D\left(x\right)$ of the real data $x$ as much as possible. When training $G$, the goal is to make $D(G(z))$ as much as possible. From \eqref{GAN}, we can calculate the optimal discriminator as \eqref{eq1}. As can be seen from \eqref{eq1} below, when $P_{data}\left(x\right)=P_z\left(z\right)$, it means that $D$ cannot distinguish whether the sample is true or false, $D$ and $G$ reach the Nash Equilibrium, and the discriminator output is 0.5.

\begin{equation}
	\begin{split}
		\label{eq1}
		D(x)=\frac{P_{data}(x)}{P_{data}(x)+P_z(z)}
	\end{split}
\end{equation}

\subsection{CGAN}
\begin{figure}[htbp]
	\centerline{\includegraphics[width=8cm, height=6cm]{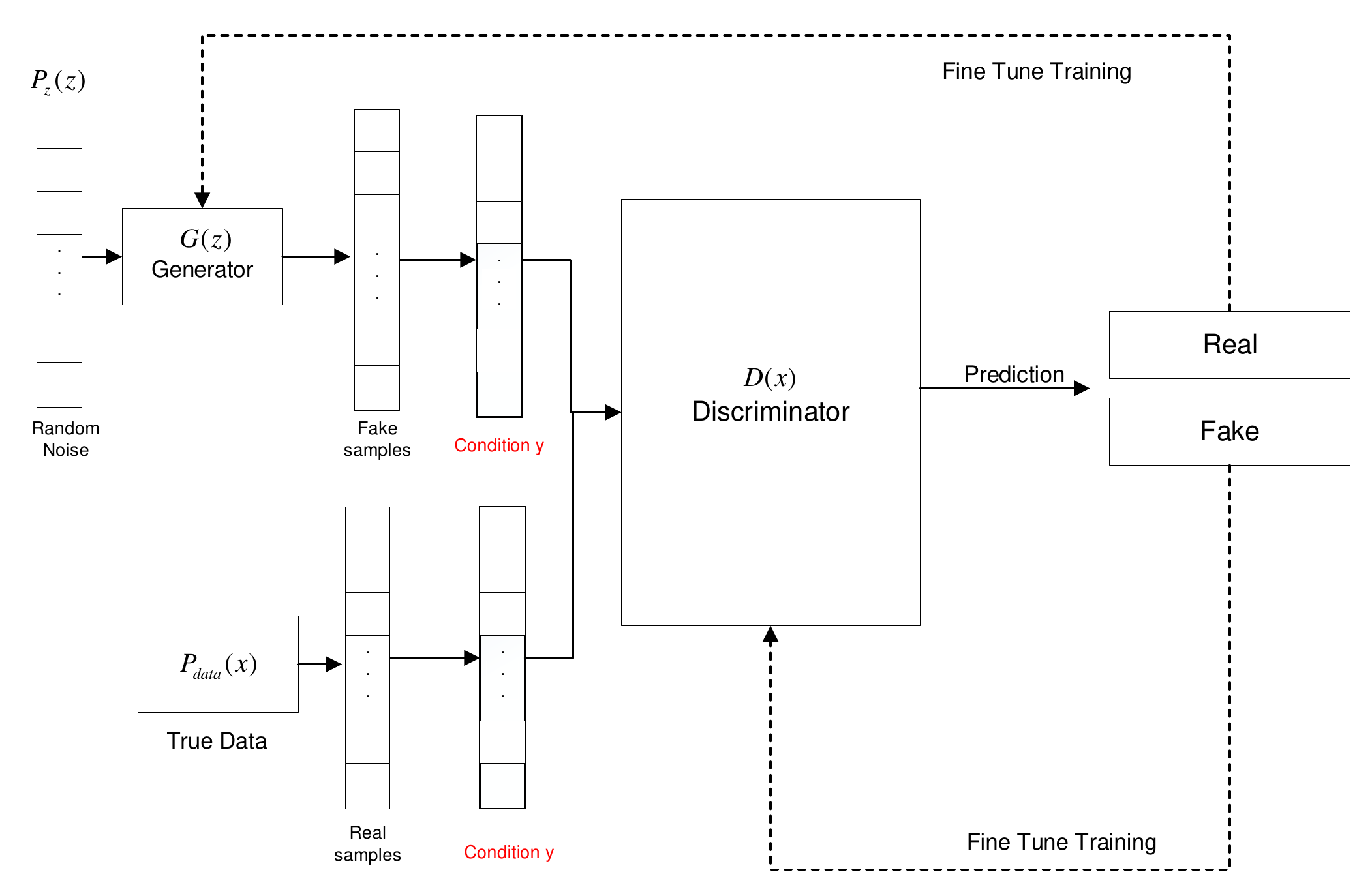}}
	\caption{The network structure of CGAN.}
	\label{fig_cgan}
\end{figure}

In GAN, there is no control over modes of the data to be generated. The conditional GAN changes that by adding the constraint condition $y$ as an additional parameter to the generator $G$ and hopes that the corresponding images are generated. For example, in MNIST, the digit generated by GAN may be either any digit from 0-9 instead of specified one or always output the same digit. Fig.~\ref{fig_cgan} shows the network structure of CGAN.


The principle, structure and training process of CGAN are similar to GAN. The cost function is slightly different as is shown in \eqref{CGAN}:

\begin{equation}
	\begin{split}
		\label{CGAN}
		\mathop {\min }\limits_{\rm{G}} \mathop {\max }\limits_D V(D,G) = \mathbb{E}_{x\sim{p_{data}}(x)}[\log D(x|y)] \\
		+ {\mathbb{E}_{z\sim{p_z}(z)}}[\log (1 - D(G(z|y)))]
	\end{split}
\end{equation}

As shown in Fig.~\ref{fig_cgan}, CGAN's training process includes the following steps:
\begin{itemize}
	\item Sampling the real data to obtain $P_{data}(x)$, obtaining the label $y$ corresponding to the sampling data $P_{data}(x)$, feed $P_{data}(x)$ and $y$ into the discriminator $D$, then updating the parameters according to the output results;
	\item Generating random noise $P_z(z)$, which is then fed into generator $G$ together with label $y$ in the above step, and $G$ generates synthesized data.
	\item Feeding the synthesized data and the label $y$ generated in the above step into the discriminator $D$, and $G$ will optimize the parameters according to the output result of $D$.
	\item Repeat the above steps until $G$ and $D$ reach the Nash equilibrium.
\end{itemize}

\section{The Methodology of PacketCGAN}\label{algorithm}
\subsection{The Workflow of PacketCGAN Based Encrypted Traffic Classification}
\begin{figure}[ht!]
	\centering\includegraphics[width=3.3 in]{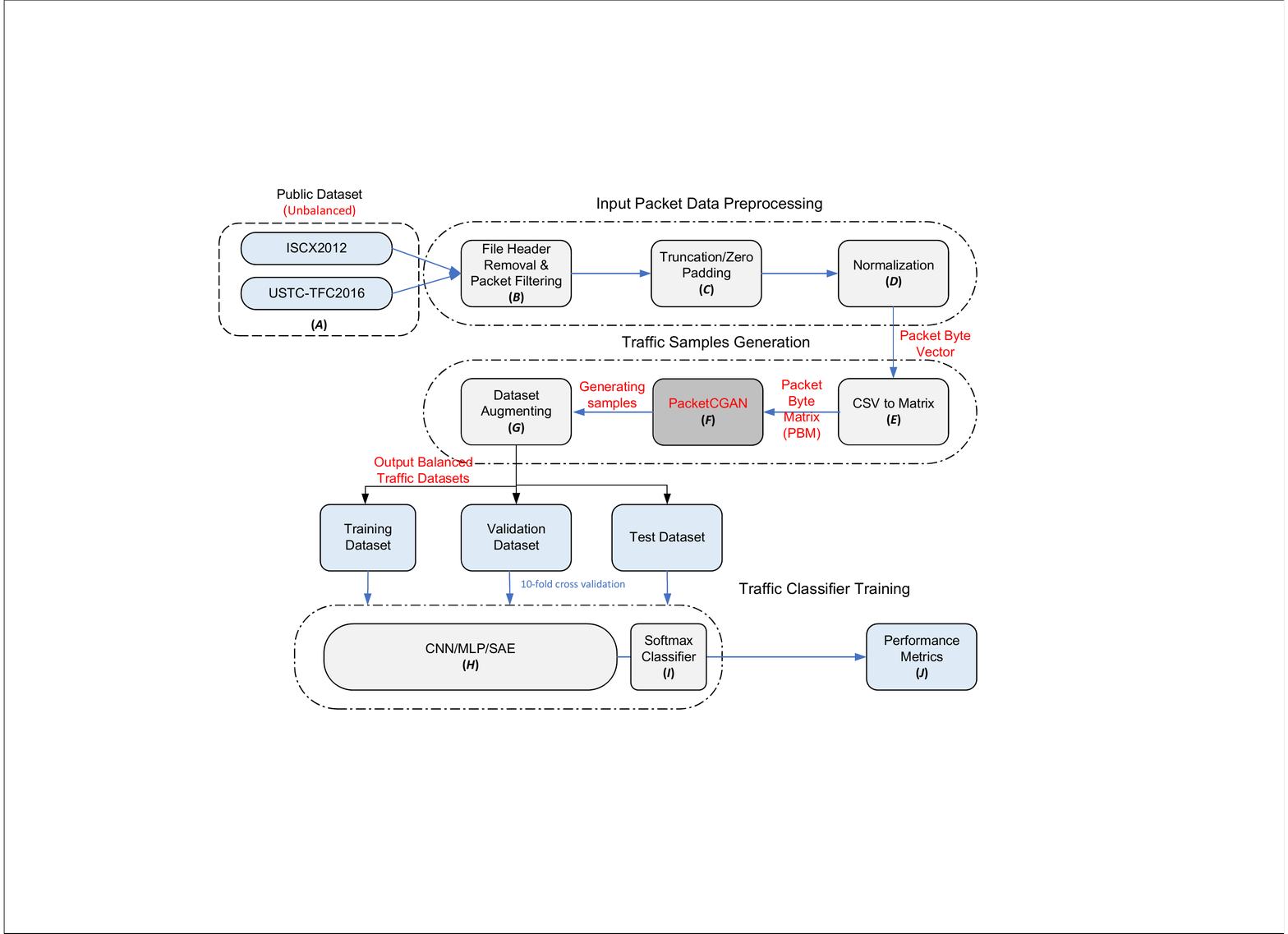} 
	\caption{The workflow of PacketCGAN.}\label{fig:workflow} 
\end{figure}

The workflow of PacketCGAN based encrypted traffic classification is shown in Fig.~\ref{fig:workflow}, in which there are three phases: packet data preprocessing, class balancing using PacketCGAN and traffic classifier training.

\subsubsection{Phase 1}\label{data_preprocess}
Generally speaking, packet data preprocessing is the fundermental task of DL modelling. As we all know, the captured network traffic data is often saved in PCAP or PCAPNG~\cite{pcap} format, in which  packet bytes of traffic data are saved in hexadecimal. The bytes converted to decimal range from 0 to 255, which resemble the pixel range of the single-channel grayscale image. Apparently, traffic classification can refer to DL methods of image recognition. However, packets from PCAP files can not directly be used for model training and need to be pre-processed. Two public datasets used in this paper are ISCX2012~\cite{ISCX} and USTC-TFC2016~\cite{Wang2D-CNN} as shown in (A) of Fig.~\ref{fig:workflow}, which are both composed of PCAP files. We will describe the two datasets in Section.~\ref{exp:dataset}.

Packet data preprocessing consists of three steps as follows:
\begin{itemize}
	\item File header removal and packet filtering as shown in (B) of Fig.~\ref{fig:workflow}. The first 24 bytes of the PCAP file header need to be removed which only contains file information and does not help with traffic classification. Filtering useless packets in PCAP file for classification is also needed, such as APR and DHCP packets etc., which is irrelevant to specific traffic types.
	\item Truncation or zero padding as shown in (C) of Fig.~\ref{fig:workflow}. During the training of DL model, the input of which needs to be converted into a vector or matrix, which requires the length of the data input to be fixed. Nevertheless, the length of captured packet in dataset tends to be different, so it is necessary to truncate the long packets or padding those short ones based on a fixed and predefined length value~\cite{r8}, which is 1480 in this paper. 
	\item Data normalization as shown in (D) of Fig.~\ref{fig:workflow}. Normalizing the pre-processed traffic data to the range [0,1] can improve the accuracy and enhance convergence speed of the model training. 
\end{itemize}

After all above mentioned operations, each processed packet will be formated to a Packet Byte Vector (PBV)~\cite{Datanet} which can be stored in csv format.

\subsubsection{Phase 2}
Data balancing is the most critical problem in this phase. It is essential to read the vectors of PBV from csv files to build the matrix of PBM as the input of PacketCGAN as shown in (E) of Fig.~\ref{fig:workflow}, which will be illustrated in detail in Section.~\ref{algorithm_description}. Synthesized packet samples will be generated by PacketCGAN in PBM format and then combined with the samples of original dataset to build the new balanced traffic dataset as shown in (E) and (F) of Fig.~\ref{fig:workflow}.

\subsubsection{Phase 3: Traffic Classifier Training}
In this paper, we adopted five DL traffic classifiers  to evaluate the performance of datasets augmented by PacketCGAN, which are MLP/CNN/SAE from ~\cite{Datanet} and 1D-CNN/2D-CNN from~\cite{Wang-1D-CNN,Wang2D-CNN}, respectively. The new balanced traffic datasets will always be devided into three parts: training, testing and validation datasets. Finally, classifier related performance metrics will be evaluated to indirectly verify the quality of datasets generated by different data augmentation methods as shown in (H,I,J) of Fig.~\ref{fig:workflow}. 

\subsection{The Model Architecture and Algorithm Description of PacketCGAN}\label{algorithm_description}
\subsubsection{The Description of Model Architecture}
\begin{figure}[ht!]
	\centering\includegraphics[width=3.3 in]{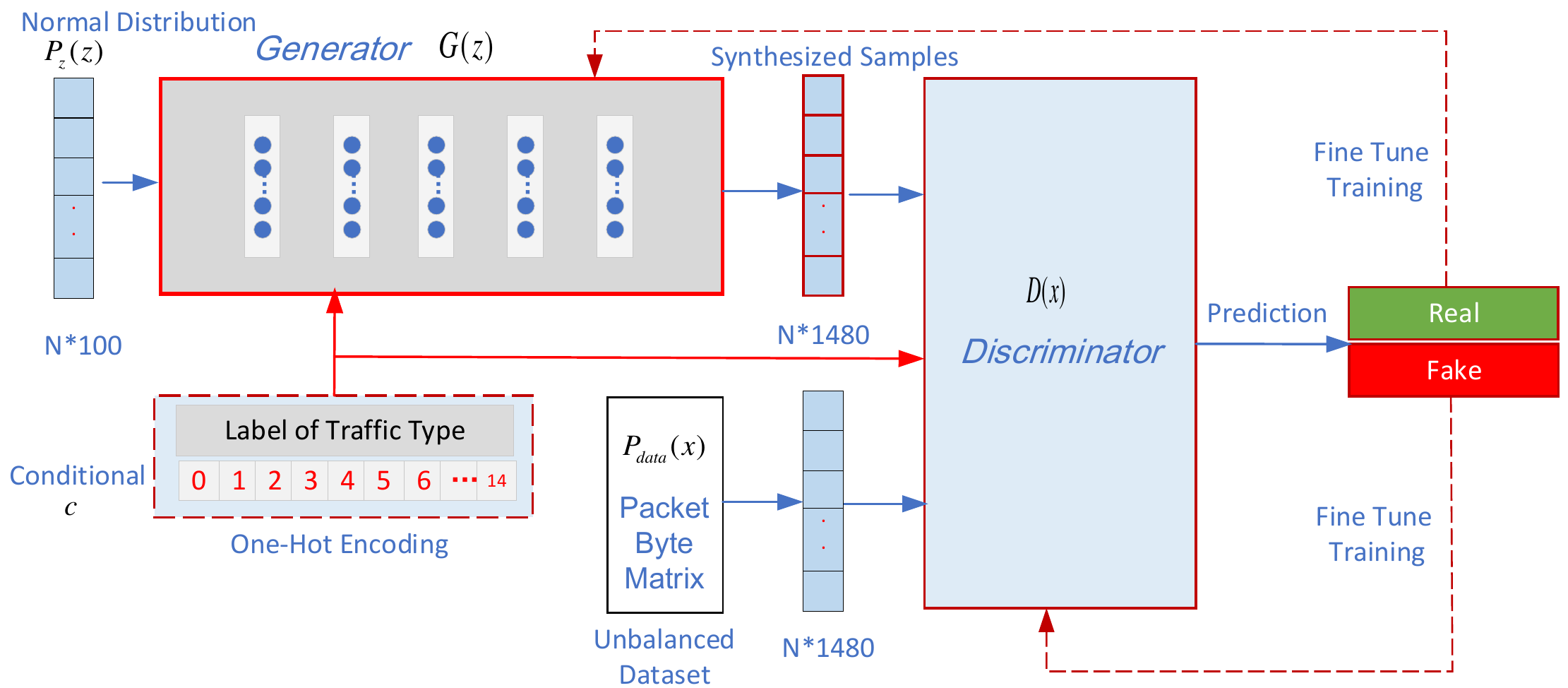} 
	\caption{The Architecture of PacketCGAN.}\label{fig:pcgan_model} 
\end{figure}

The model of our proposed PacketCGAN is shown in Fig.~\ref{fig:pcgan_model}, in which both $G$ and $D$ adopted MLP as the basic architecture. Random noise $P_z(z)$ as the input of model with Normal Distribution are N*100-dimentional vectors, which will be fed into $G$ with $G(z)$ distribution. Different from vanilla GAN, the label of traffic/applications types with one-hot encoding will also be fed into our PacketCGAN as conditional $c$. The synthesized samples as the output of $G$ are N*1480 vectors, which will be compose of the matrix of PBM and fed into $D$ with $D(x)$ distribution eventually. Meanwhile, PBM from original unbalanced datasets with $P_{data}(x)$ distribution will be fed into $D$ with the same 1480-dimentional to compete to reach Nash equilibrium with fine-tune training.

\subsubsection{The Description of Model Algorithm}\label{model_algorithm}
In general, PacketCGAN's training mainly includes the following three steps: the training of discriminator $D$, generator $G$ and fine-tune optimization. During the training of $D$ in step 1, one needs to fix the parameters of $G$ while training $D$. We adopt the simple MLP to build the $D$ model, in which there are three layers: the input layer, a hidden layer and the output layer.  The data for the input layer is derived from the N*1480 matrix of  PBM aforementioned in Section~\ref{data_preprocess} with the distribution of $P_{data}(x)$. The hidden layer consists of 128 neurons with the notation of $D_{h1}$.

\begin{equation}
	\begin{split}
		\label{d_input}
		D_{h1}=ReLU(input_D\cdot W^D_1+b^D_1)
	\end{split}
\end{equation}

The $input_D$ in  \eqref{d_input} is composed of the real data $x$ with $P_{data}(x)$ and the traffic/applications types label $y$ in one-hot encoding, such as 0 for SSH, 1 for YouTube. The hidden layer $D_{h1}$ adopts $ReLU$ as the activation function. The output layer has only one neuron referring to real or fake sample as following Equation:~\ref{d_output}:

\begin{equation}
	\begin{split}
		\label{d_output}
		D_{out}=D_{h1}\cdot W^D_2+b^D_2
	\end{split}
\end{equation}

The model uses the Adam optimization algorithm as the optimizer to update weights $W^D_1$, $W^D_2$ and bias $b^D_1$,  $b^D_2$. The overall training process of the discriminator is shown in Algorithm.~\ref{algorithm:discriminator}, in which $X_{\tau}=\begin{pmatrix}
x_{11}&\cdots&x_{1k}\\
\vdots&\ddots&\vdots\\
x_{m1}&\cdots&x_{mk}
\end{pmatrix}$
,$y_{\tau}=\begin{pmatrix}
y_{11}\\
\vdots\\
y_{m1}
\end{pmatrix}$
,k is 1480, which is the data dimension. $m$ is mini\_batch referring to a group of packets from dataset.

\begin{algorithm}[htbp]
	\caption{The training of Discriminator of PacketCGAN}  \label{algorithm:discriminator}
	\begin{algorithmic}[1]  
		\REQUIRE real data $X_{\tau}$ and label $y_{\tau}$ from Section ~\ref{model_algorithm}, $\tau$ is the number of iterations
		\ENSURE the probability of the output data being real or false 
		\STATE Initializing the relevant parameters, e represents the training cycle: epoches.
		\FOR {$\tau$ in $e$ }
		\FOR {each batch of $m$ input data}
		\STATE concatenate {\tiny } $X_\tau$ with $y_\tau$: 
		\STATE Compute the output using Equation.~(\ref{d_input});
		\STATE Compute the output using Equation.~(\ref{d_output});
		\STATE Output discriminating results according to Equation.~(\ref{d_output});
		\STATE Optimize the loss function
		\STATE Update weights and bias;
		\ENDFOR \\ 
		\ENDFOR \\ 
	\end{algorithmic}  
\end{algorithm}  

Step 2 is to train the generator $G$. The parameters of $D$ need to be fixed like step 1. The input layer of $G$ consists of 115 neurons, which includes stochastic noise $P_z(z)$ with 100 neurons and sample label $y$ with 15 neurons of applications types. The hidden layer also has 128 neurons to update the weight and bias of the input data:

\begin{equation}
	\begin{split}
		\label{g_input}
		G_{h1}=ReLU(input_G\cdot W^G_1+b^G_1)
	\end{split}
\end{equation}

The 'input' in \eqref{g_output} is made up of random noise $P_z(z)$ and the sample label $y$ with $ReLu$ as the activation function. The output layer contains 1480 neurons with the same dimension of the $P_{data}(x)$. The computed result will be activated by the sigmoid function for discrimination of real or fake.

\begin{equation}
	\begin{split}
		\label{g_output}
		G\_out=sigmoid(G_{h1}\cdot W^D_2+b^D_2)
	\end{split}
\end{equation}

The optimization function is the same as the discriminator. The generator training algorithm is shown in Algorithm.~\ref{algorithm:generator}. The sample label $y$ in Algorithm.~\ref{algorithm:generator} should be consistent with $y$ in Algorithm.~\ref{algorithm:discriminator}.
$Z_{\tau}=\begin{pmatrix}
z_{11}&\cdots&z_{1n}\\
\vdots&\ddots&\vdots\\
z_{m1}&\cdots&z_{mn}
\end{pmatrix}$,$n=100$.

\begin{algorithm}[htbp]
	\caption{The training of Generator of PacketCGAN}  \label{algorithm:generator}
	\begin{algorithmic}[1]  
		\REQUIRE random noise $Z_{\tau}$, and label $y_{\tau}$, $\tau$ is the number of iterations
		\ENSURE synthesized data $g_{\tau}$
		\STATE Set the relevant initialization parameters, $e$ represents the training cycle: epoches.
		\FOR {$\tau$ in $e$ }
		\FOR {each batch of $m$ input data}
		\STATE concatenant $Z_\tau$ with $y_\tau$: 
		\STATE Compute the output using Equation.~(\ref{g_input});
		\STATE Compute the output using Equation.~(\ref{g_output});
		\STATE Output generation data according to Equation.~(\ref{g_output});
		\STATE Optimal the loss function
		\STATE Update weights and bias;
		\ENDFOR \\ 
		\ENDFOR \\ 
	\end{algorithmic}  
\end{algorithm}  

\section{Evaluation and Experimental results}\label{exp:results}
\subsection {Experimental Settings}\label{exp:setting}
\subsubsection{Dataset for Evaluation}\label{exp:dataset}
Two public datasets used in this paper are ISCX2012~\cite{ISCX} and USTC-TFC2016~\cite{Wang2D-CNN}, which are both composed of PCAP files. The full name of 'ISCX2012' dataset is 'ISCX VPN\-nonVPN traffic dataset', which contains many encryption applications or protocols such as HTTPS, SFTP, Facebook, Hangouts, etc. In addition to these applications encapsulated in regular session, some VPN tunnel's applications were also captured in this dataset. 15 applications were chosen to build the original dataset as shown in Table~\ref{tab:Desc_Samples_ISCX}. Apparently, the original chosen dataset happens to the problem of class imbalance, in which the majority class such as Netflix accounting for 25.13\%, while the minor class ICQ account for only 2.05\%. 10 applications were chosen from USTC-TFC2016 as shown in Table~\ref{tab:Desc_Samples_USTC}. Balanced dataset made by different generative methods are all 10000 samples per applications.

\begin{table}[htbp]	
	\centering  
	\fontsize{6.5}{8}\selectfont  
	\begin{threeparttable}  
		\caption{Description of the chosen datasets from ISCX.}  \label{tab:Desc_Samples_ISCX}  
		\begin{tabular}{l|c|cc|cc|}  
			\toprule  
			\multirow{2}{*}{\textbf{Application}}&
			\multirow{2}{*}{\textbf{Security}}&
			\multicolumn{2}{c}{\textbf{Unbalanced dataset}}&\multicolumn{2}{|c}{\textbf{Balanced  dataset}} \cr  
			\cmidrule(lr){3-4} \cmidrule(lr){5-6}  
			&\textbf{Protocol} & \textbf{Quantity} & \textbf{Percentage} & \textbf{Quantity} & \textbf{Percentage}\cr  
			
			\midrule  
			AIM				&HTTPS	 	   &4869&2.356\%	  &10000&6.67\%\cr  
			Email-Client&SSL			  &4417&2.137\%		 &10000&6.67\%\cr  
			Facebook	&HTTPS	 	   &5527&2.674\%	  &10000&6.67\%\cr  
			Gmail		   &HTTPS		  &7329&3.546\%		 &10000&6.67\%\cr  
			Hangout		&HTTPS	 	   &7587&3.671\%	  &10000&6.67\%\cr  
			ICQ				&HTTPS	 	   &4243&2.053\%	  &10000&6.67\%\cr  
			Netflix			&HTTPS		   &51932&25.126\%	&10000&6.67\%\cr  
			SCP				&SSH	  		 &15390&7.446\%	   &10000&6.67\%\cr  
			SFTP			&SSH	 		 &4729&2.287\%	    &10000&6.67\%\cr  
			Skype		   &proprietary	 &4607&2.229\%	    &10000&6.67\%\cr  
			Spotify		   &proprietary	 &14442&6.987\%	   &10000&6.67\%\cr  
			torTwitter	  &proprietary	&14654&7.089\%	  &10000&6.67\%\cr  
			Vimeo		  &HTTPS		  &18755&9.074\%	&10000&6.67\%\cr  
			voipbuster	&proprietary   &35469&17.161\%	&10000&6.67\%\cr  
			Youtube		&HTTPS			 &12738&6.163\%	   &10000&6.67\%\cr  							
			\midrule
			
			TOTAL&  &{\bf 206688}&{\bf 100\%}&{\bf 150000}&{\bf 100\%}
		\end{tabular}  
	\end{threeparttable}  
\end{table}

\begin{table}[htbp]	
	\centering  
	\fontsize{6.5}{8}\selectfont  
	\begin{threeparttable}  
		\caption{Description of the chosen datasets from USTC-TFC2016.}  \label{tab:Desc_Samples_USTC}  
		\begin{tabular}{l|c|cc|cc|}  
			\toprule  
			\multirow{2}{*}{\textbf{Application}}&
			\multirow{2}{*}{\textbf{Security}}&
			\multicolumn{2}{c}{\textbf{Unbalanced dataset}}&\multicolumn{2}{|c}{\textbf{Balanced  dataset}} \cr  
			\cmidrule(lr){3-4} \cmidrule(lr){5-6}  
			&\textbf{Protocol} & \textbf{Quantity} & \textbf{Percentage} & \textbf{Quantity} & \textbf{Percentage}\cr  
			
			\midrule  
			BitTorrent				&SSL	 	   &7535&8.10\%	  &10000&10.00\%\cr  
			Facetime&SSL			  &2990&3.22\%		 &10000&10.00\%\cr  
			FTP		   &FTP		  &11506&12.37\%		 &10000&10.00\%\cr  
			Gmail		&HTTPS	 	   &11477&12.34\%	  &10000&10.00\%\cr  
			MySQL				&MySQL	 	   &11385&12.24\%	  &10000&10.00\%\cr  
			Outlook			&SSL		   &7467&8.03\%	&10000&10.00\%\cr  
			Skype		   &proprietary	 &6028&6.48\%	    &10000&10.00\%\cr  
			SMB		   &SMB	 &11543&12.41\%	   &10000&10.00\%\cr  
			Weibo	  &HTTPS	&11510&12.38\%	  &10000&10.00\%\cr  
			WorldOfWarcraft		  &SSL		  &11559&12.43\%	&10000&10.00\%\cr  
			\midrule
			
			TOTAL&  &{\bf 93000}&{\bf 100\%}&{\bf 100000}&{\bf 100\%}
		\end{tabular}  
	\end{threeparttable}  
\end{table}

\subsubsection{Configurations of the Computing Platform}\label{exp:config}
The experimental environmental parameters of this paper are shown in Table~\ref{tab:parameters}. The performance evaluations are conducted using a Dell R730 server with an Intel I7-7600U CPU 2.8 GHz, 16 GB RAM and an external GPU (Nvidia GeForce GTX 1050TI). The software platform for deep learning is built on Keras library with Tensorflow (GPU-based version 1.13.1) as the back-end support.

\subsubsection{Description of deep learning based network traffic classifier}\label{exp:classifier}
To verify the feasibility and evaluate the performance of the PacketCGAN algorithm, we adopted three classical DL models to classify the traffic over the datasets synthesized by different generative methods, which is MLP, CNN and SAE, respectively. The detail of these three models can be find in our previous works~\cite{Datanet}.

\linespread{1.5}
\begin{table}[htbp]
	\caption{Experimental Environment Parameters}
	\begin{center}
		\begin{tabular}{c c}
			\hline
			\textbf{Category}&{\textbf{Parameters}} \\
			\hline
			GPU & Nvidia GPU(GeForce GTX 1050Ti)  \\
			Operating System & Win 10 \\
			Deep learning platform & TensorFlow 1.13.1 + Keras 1.0.7\\
			CUDA Version & 9.0\\
			CuDNN Version & 7.6.0\\
			\hline
		\end{tabular}
		\label{tab:parameters}
	\end{center}
\end{table}

\subsubsection{Performance Metrics for Classification}

The performance metrics used for evaluations of network traffic classifiers are \emph{Precision}, \emph{Recall} and \emph{$F_1$ score}.
\begin{itemize}
	
	\item \textbf{\emph{Precision}}: precision $r_{p}$ is the ratio of \emph{true positives} $n_T^P$ over the sum of $n_T^P$ and \emph{false positives} $n_F^P$. In the proposed classification methods, precision is the percentage of packets that are properly attributed to the targeted application.
	\begin{equation}
		r_p = \frac{n_T^P}{n_T^P+n_F^P}.
	\end{equation}
	
	\item \textbf{\emph{Recall}}: recall $r_c$ is the ratio of $n_T^P$ over the sum of $n_T^P$ and \emph{false negatives} $n_F^N$ or the percentage of packets in an application class that are correctly identified.
	\begin{equation}
		r_c = \frac{n_T^P}{n_T^P+n_F^N}.
	\end{equation}
	
	\item \textbf{\emph{$F_1$-score}}: the $F_1$ score $r_f$ is a widely-used metric in information retrieval and classification that considers both precision and recall as follows:
	\begin{equation}
		r_f = \frac{2 r_p \cdot r_c}{r_p + r_c}.
	\end{equation}
	
\end{itemize}

\subsection {Data Augmenting using PacketCGAN}
\subsubsection {Data Augmenting Methods handling the problem of class imbalance}
In this paper, four methods were used to address the problem of the unbalanced dataset for the purpose of comparism. The first one is random oversampling method (ROS) mentioned in Section~\ref{subsec_imbalance} . The essence of this method is to randomly copy some samples of minor class to supplement the dataset. Since the simple method is only a copy of the original data, it will lead to some wrong features be learned by the model easily and even over-fitting. The second method is SMOTE in which minority samples are generated by synthetic samples rather than copies. The third one is vanilla GAN with a generator and discriminator based on MLP. The last method is to generate minor sample data using our PacketCGAN generator. The dataset balanced by different methods is described in ~\ref{exp:dataset}.

\subsubsection {The Architecture of PacketCGAN}
In our experiments, a fully connected MLP network was adopted to design generators and discriminators, as shown in Table~\ref{tab:generator} and ~\ref{tab:discriminator}. The generator's input is a 100-dimensional vector generated by random Gaussian noise with additional 15 applications types for ISCX2012 or 10 for USTC-TFC2016 datasets. The next three hidden layers have 128, 256, 512 neurons, and 1480 output neurons with additional 15 for ISCX2012 or 10 for USTC-TFC2016 datasets. Therefore, input of discriminator network is 1495 or 1490 dimension vectors mixed by real traffic data or generated data. Three hidden layers and output layers are both activated by LeakyReLU.

\linespread{1.5}
\begin{table}[htbp]
	\caption{The Architecture of Generator Network of PacketCGAN}
	\fontsize{6.5}{8}\selectfont  
	\begin{center}
		\begin{tabular}{c| c| c}
			\hline
			\textbf{Layer(type)}&{\textbf{Output Shape (ISCX)}} &{\textbf{Output Shape (USTC)}} \\
			\hline
			dense\_1(Dense) 	& 115 	& 110  \\
			LeakyReLU 			& 115  	& 110   \\
			dense\_2(Dense) 	& 256 	& 256	\\
			LeakyReLU 			& 256	& 256    \\
			dense\_3(Dense) 	& 512 	& 512   \\
			LeakyReLU 			& 512	& 512   \\
			dense\_4(Dense)		 & 1495 & 1490	\\
			LeakyReLU 			& 1495		& 1490 \\	
			
			\hline
		\end{tabular}
		\label{tab:generator}
	\end{center}
\end{table}

\linespread{1.5}
\begin{table}[htbp]
	\caption{The Architecture of Discriminator Network of PacketCGAN}
	\fontsize{6.5}{8}\selectfont  
	\begin{center}
		\begin{tabular}{c| c| c}
			\hline
			\textbf{Layer(type)}&{\textbf{Output Shape (ISCX)}} &{\textbf{Output Shape (USTC)}} \\
			\hline
			dense\_6(Dense) 	& 1495 	& 1490  \\
			LeakyReLU 			& 1495  	& 1490   \\
			Dropout 				& 1495  	& 1490   \\
			dense\_7(Dense) 	& 512 	& 512	\\
			LeakyReLU 			& 512	& 512    \\
			Dropout 				& 512  	& 512   \\
			dense\_8(Dense) 	& 256 	& 256   \\
			LeakyReLU 			& 256	& 256   \\
			Dropout 				& 256  	& 256   \\
			dense\_9(Dense)		 & 128 	& 128	\\
			LeakyReLU 			& 128		& 128 \\
			Dropout 				& 128  	& 128   \\	
			dense\_10(Dense) 	& 1  	& 1   \\
			
			\hline
		\end{tabular}
		\label{tab:discriminator}
	\end{center}
\end{table}

\subsubsection {The Training of PacketCGAN}
Table~\ref{training_parameters} shows the parameters of the optimizer, loss function, epoches, and mini\_batch used in the deep training model training process.

\linespread{1.5}
\begin{table}[htbp]
	\fontsize{6.5}{8}
	\caption{Training Parameters of Packet CGAN}
	\begin{center}
		\begin{tabular}{c c c }
			\hline
			\textbf{Training Parameters}&{\textbf{Generator}}&{\textbf{Discriminator}} \\
			\hline
			Optimizer & Adma & Adma   \\
			Loss Function & Cross-Entropy & Cross-Entropy  \\
			Epoches & 200000 & 200000 \\
			Mini\_batch & 64 & 64 \\
			\hline
		\end{tabular}
		\label{training_parameters}
	\end{center}
\end{table}

\begin{figure}[htbp]
	\centering
	\subfigure[d\_loss for ISCX2012]{
		\begin{minipage}[t]{0.5\linewidth}
			\centering
			\includegraphics[width=1.6in]{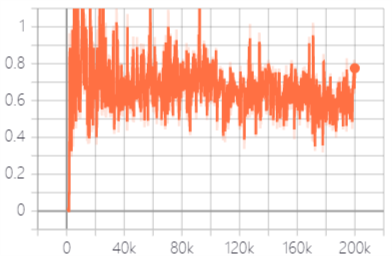}
		\end{minipage}%
	}%
	\subfigure[g\_loss for ISCX2012]{
		\begin{minipage}[t]{0.5\linewidth}
			\centering
			\includegraphics[width=1.7in]{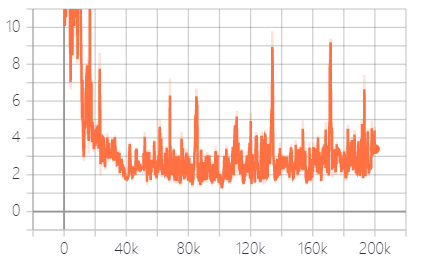}
		\end{minipage}%
	}%
	\\
	\centering
	\subfigure[d\_loss for USTC-TFC2016]{
		\begin{minipage}[t]{0.5\linewidth}
			\centering
			\includegraphics[width=1.6in]{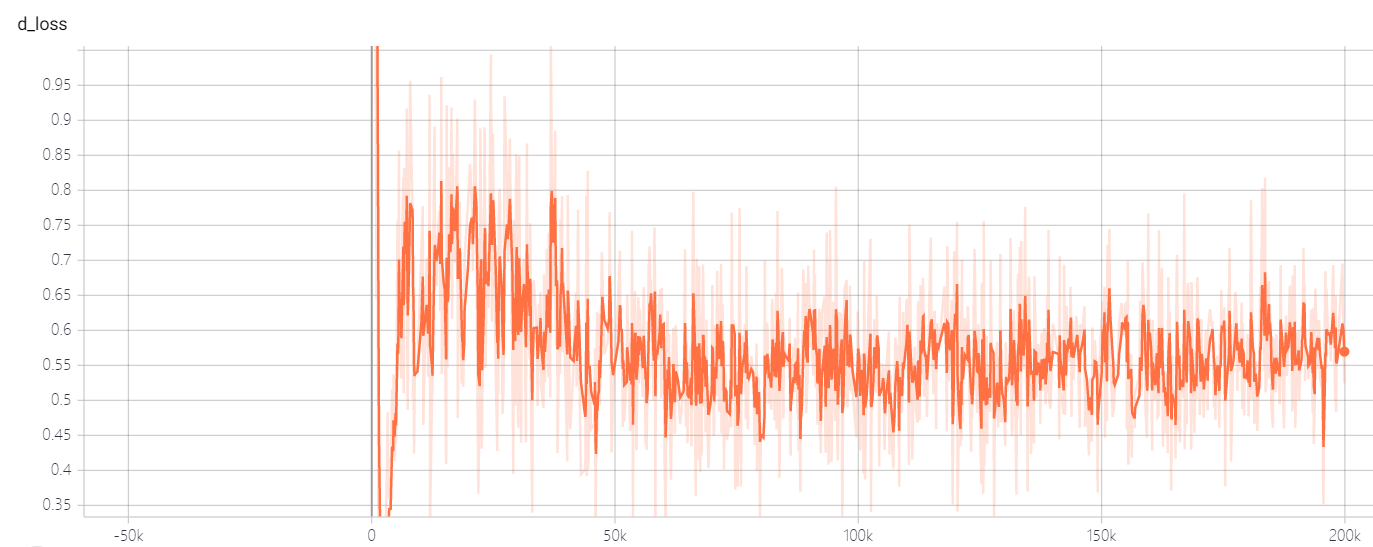}
		\end{minipage}%
	}%
	\subfigure[g\_loss for USTC-TFC2016]{
		\begin{minipage}[t]{0.5\linewidth}
			\centering
			\includegraphics[width=1.7in]{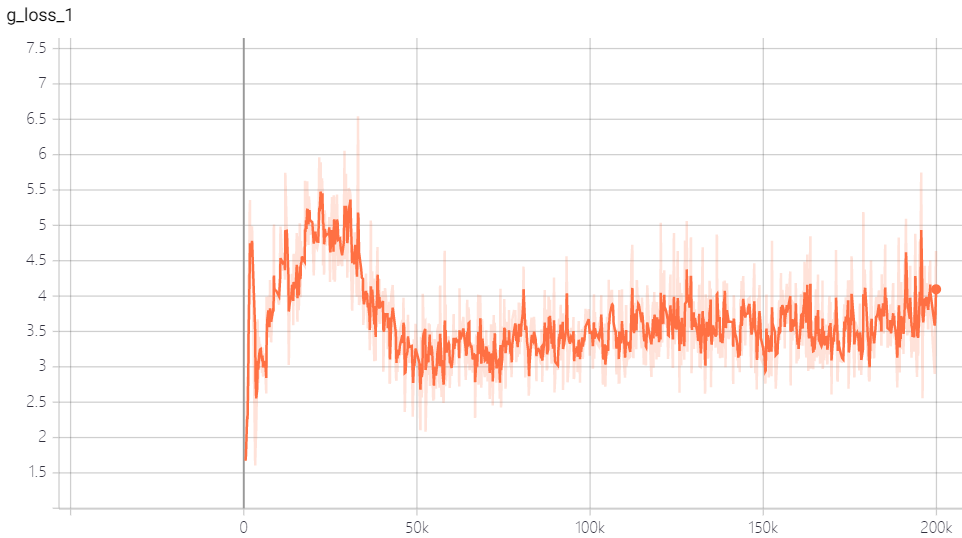}
		\end{minipage}%
	}%
	
	\caption{Loss of PacketCGAN}
	\label{cgan_loss}
\end{figure}

Figure~\ref{cgan_loss} shows the trend of loss of generator and discriminator during the PacketCGAN training process. It can be seen that PacketCGAN does not improve the unstability characteristics of GAN. The loss of $G$ and $D$ always fluctuates within a range instead of convergence. Nevertheless, it is a big improvement that only one CGAN model needs to be trained when using PacketCGAN compared with vanilla GAN. One can generate any samples with the additional input of the applications types as conditional $c$, while vanilla GAN can only generate one type of samples at a time, that means we can control the output of the PacketCGAN generator so that we can easily generate specified minor classes samples needed to be augmented to handle the problem of class balance.

\begin{figure*}[htb]
	\centering
	\subfigure[ROS]{
		\begin{minipage}[t]{0.25\linewidth}
			\centering
			\includegraphics[width=1.7in]{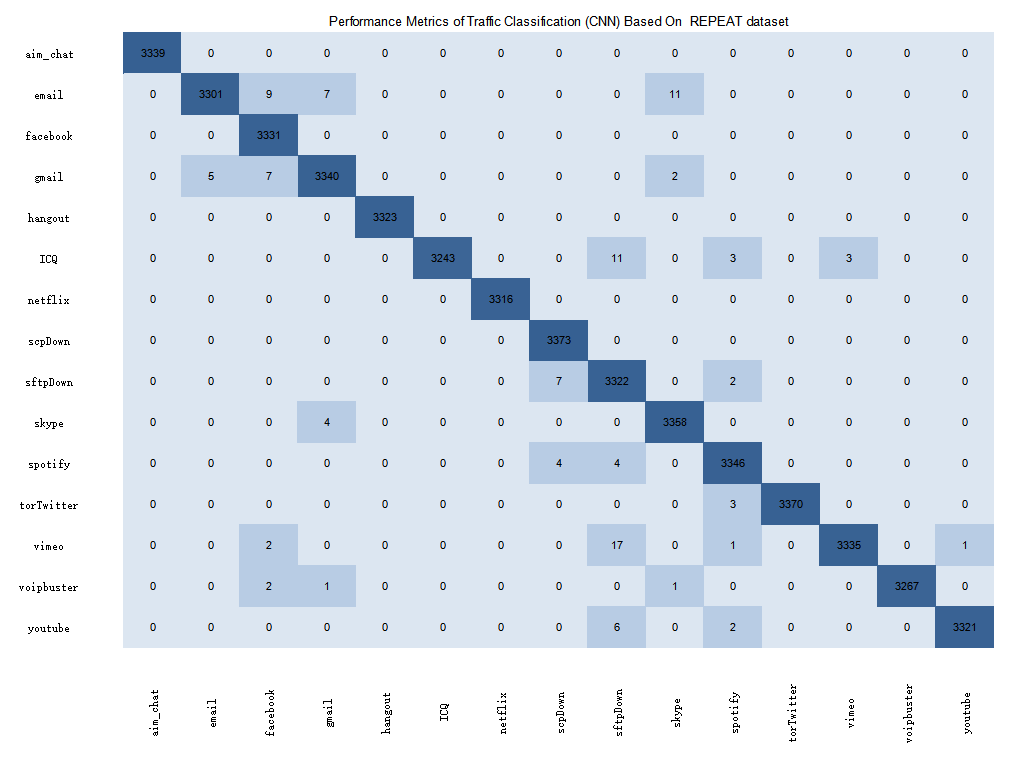}
			\label{ros_cm_iscx}
		\end{minipage}%
	}%
	\subfigure[SMOTE]{
		\begin{minipage}[t]{0.25\linewidth}
			\centering
			\includegraphics[width=1.7in]{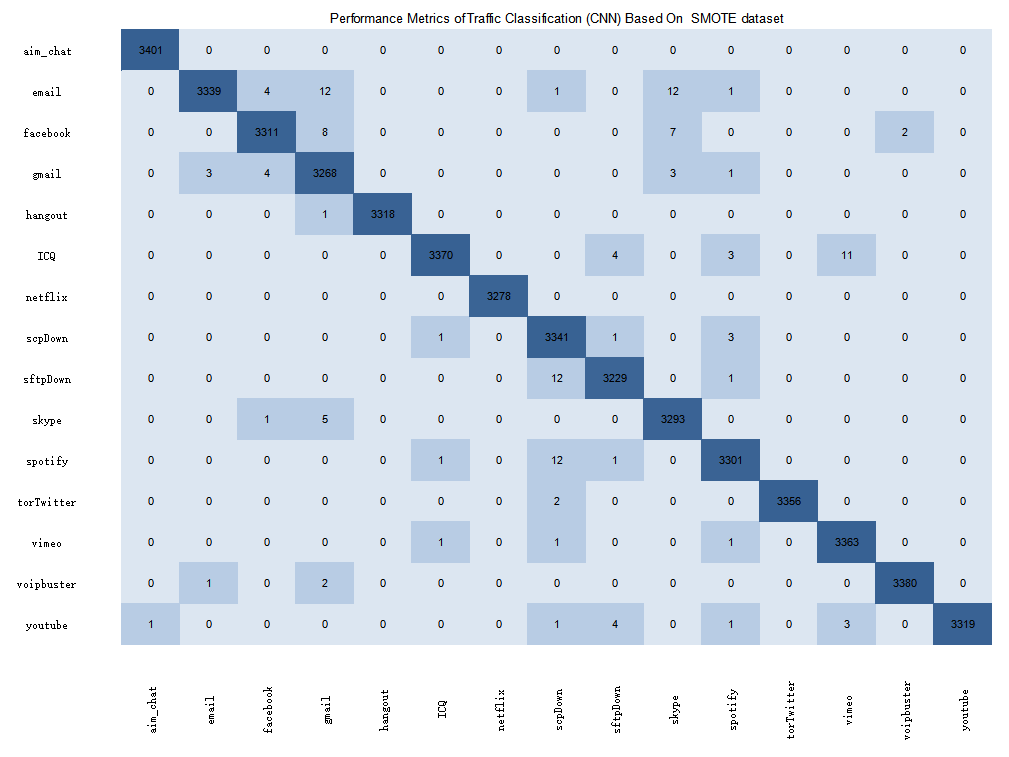}
			\label{smote_cm_iscx}
		\end{minipage}%
	}%
	\subfigure[GAN]{
		\begin{minipage}[t]{0.25\linewidth}
			\centering
			\includegraphics[width=1.7in]{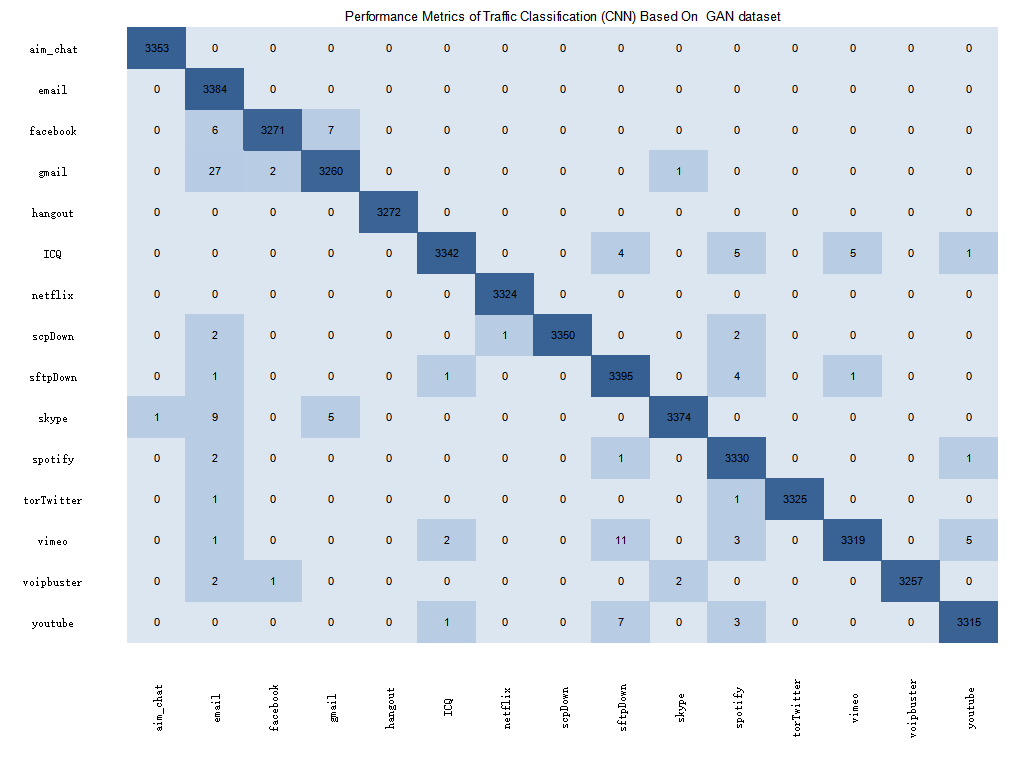}
			\label{gan_cm_iscx}
		\end{minipage}%
	}%
	\subfigure[PacketCGAN]{
		\begin{minipage}[t]{0.25\linewidth}
			\centering
			\includegraphics[width=1.7in]{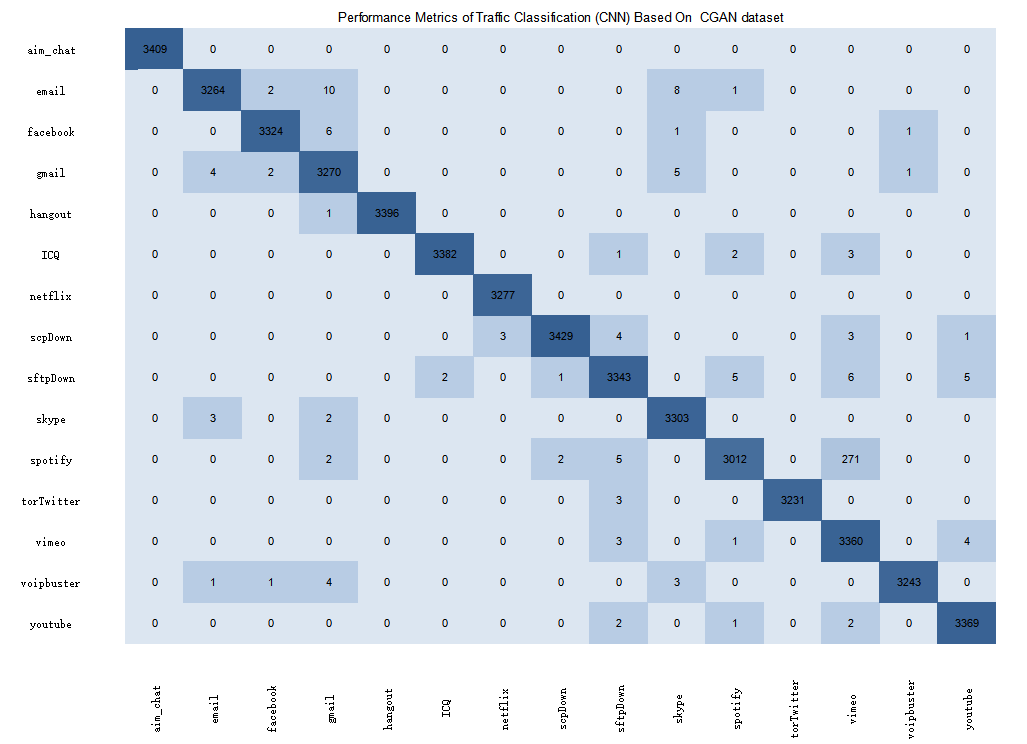}
			\label{cgan_cm_iscx}
		\end{minipage}%
	}%
	\caption{Confusion Matrices of CNN-based encrypted traffic identification method on ISCX2012 dataset}
	\label{confusion_matrices_ISCX}

	\centering
	\subfigure[ROS]{
		\begin{minipage}[t]{0.25\linewidth}
			\centering
			\includegraphics[width=1.7in]{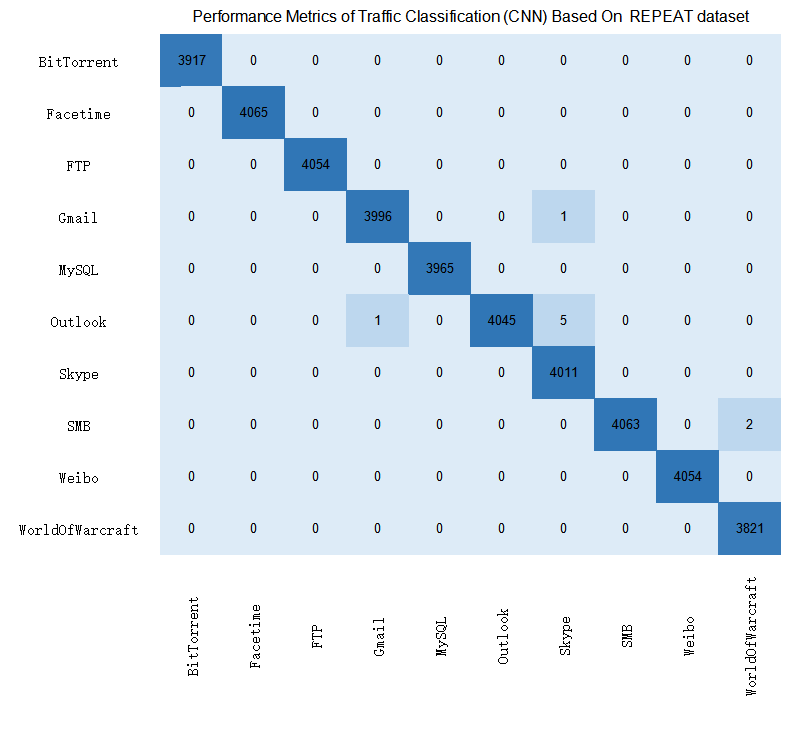}
			\label{ros_cm_ustc}
		\end{minipage}%
	}%
	\subfigure[SMOTE]{
		\begin{minipage}[t]{0.25\linewidth}
			\centering
			\includegraphics[width=1.7in]{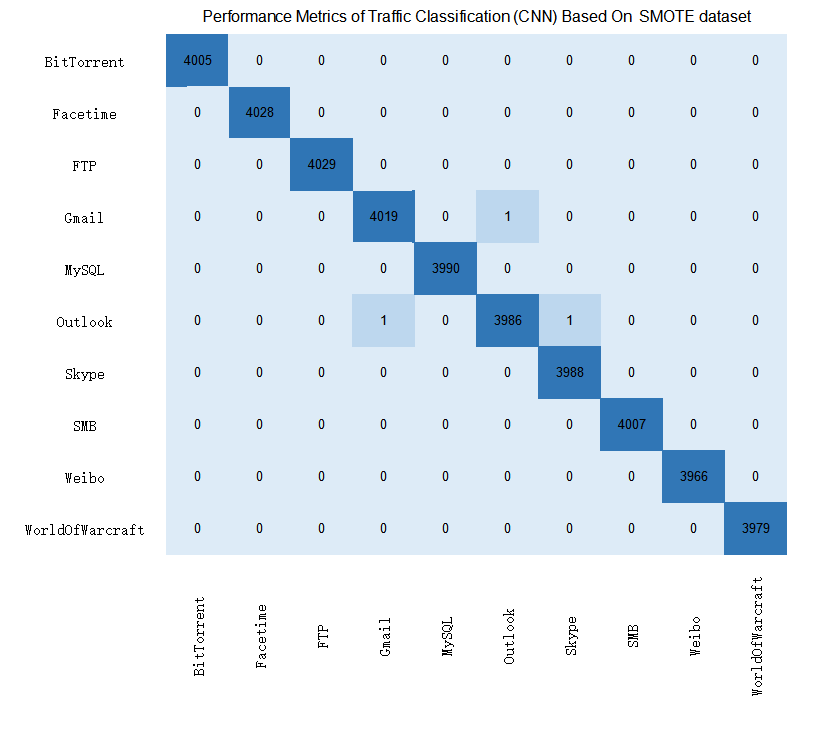}
			\label{smote_cm_ustc}
		\end{minipage}%
	}%
	\subfigure[GAN]{
		\begin{minipage}[t]{0.25\linewidth}
			\centering
			\includegraphics[width=1.7in]{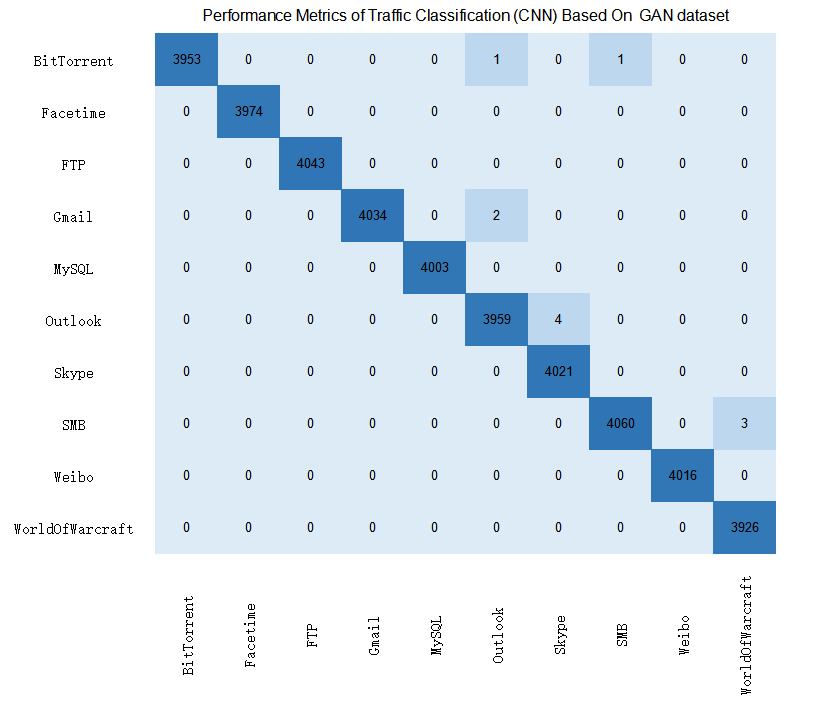}
			\label{gan_cm_ustc}
		\end{minipage}%
	}%
	\subfigure[PacketCGAN]{
		\begin{minipage}[t]{0.25\linewidth}
			\centering
			\includegraphics[width=1.7in]{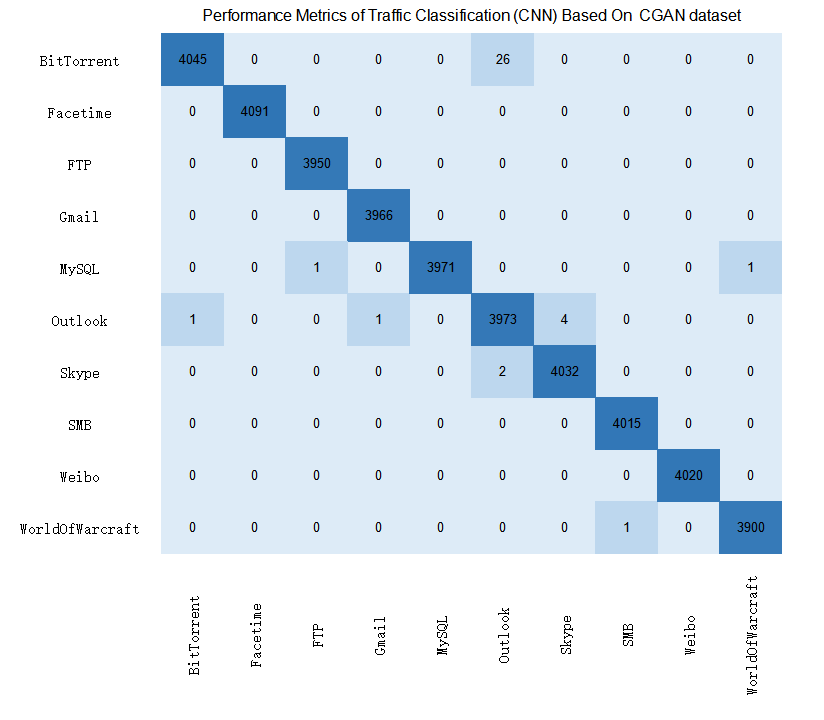}
			\label{cgan_cm_ustc}
		\end{minipage}%
	}%
	\caption{Confusion Matrices of CNN-based encrypted traffic identification method on USTC-TFC2016}
	\label{confusion_matrices_USTC}
\end{figure*}

Fig.~\ref{confusion_matrices_ISCX} and ~\ref{confusion_matrices_USTC} shows the Confusion Matrices of the CNN-based encrypted traffic identification model based on 8 datasets augmented by four different generative methods, which are ROS/SMOTE/GAN/CGAN over ISCX2012 and USTC-TFC2016, respectively. The elements on the diagonal of Confusion Matrix refer to the correct ones of classification and all the others are mis-classified. It can be clearly seen that the false positive rate of minor class in Fig~\ref{cgan_cm_iscx} and ~\ref{cgan_cm_ustc} is lower than ROS as shown in  Fig~\ref{ros_cm_iscx} and ~\ref{ros_cm_ustc}, SMOTE as shown in  Fig~\ref{smote_cm_iscx} and ~\ref{smote_cm_ustc} and GAN as shown in  Fig~\ref{gan_cm_iscx} and ~\ref{gan_cm_ustc}.

\begin{figure*}[htb]
	\centering
	\subfigure[ROS]{
		\begin{minipage}[t]{0.25\linewidth}
			\centering
			\includegraphics[width=1.7in]{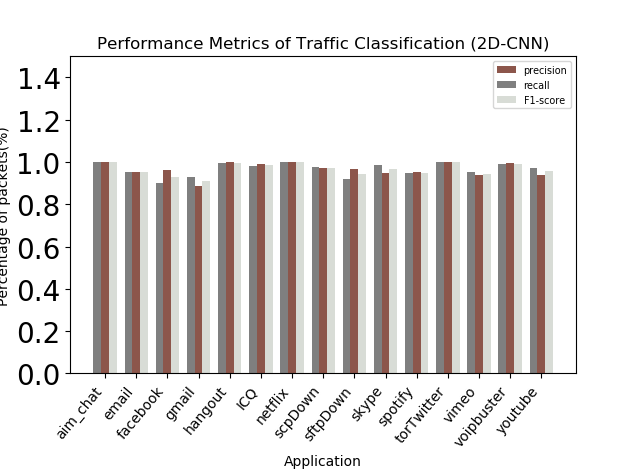}
		\end{minipage}%
	}%
	\subfigure[SMOTE]{
		\begin{minipage}[t]{0.25\linewidth}
			\centering
			\includegraphics[width=1.7in]{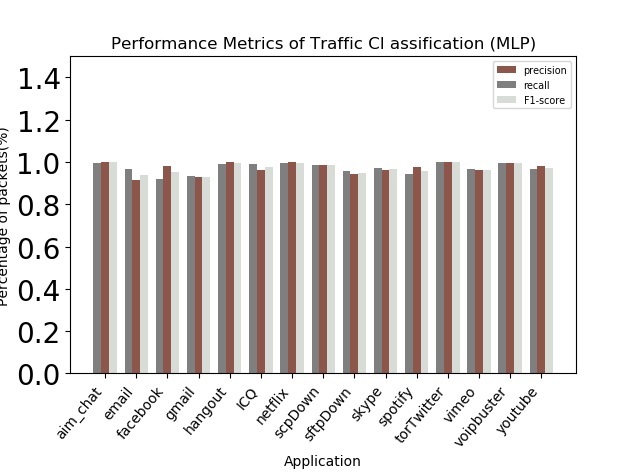}
		\end{minipage}%
	}%
	\subfigure[GAN]{
		\begin{minipage}[t]{0.25\linewidth}
			\centering
			\includegraphics[width=1.7in]{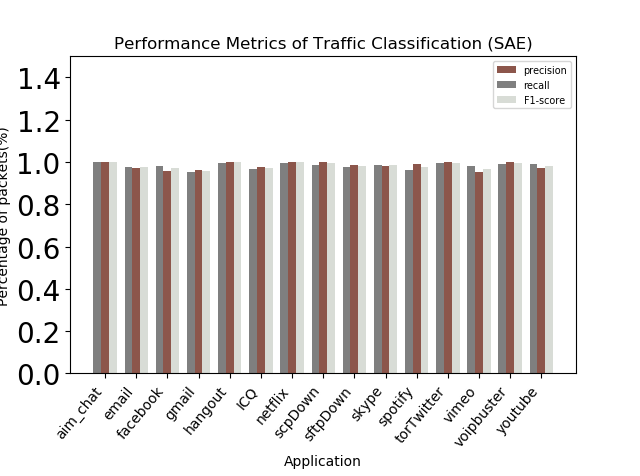}
		\end{minipage}%
	}%
	\subfigure[PacketCGAN]{
		\begin{minipage}[t]{0.25\linewidth}
			\centering
			\includegraphics[width=1.7in]{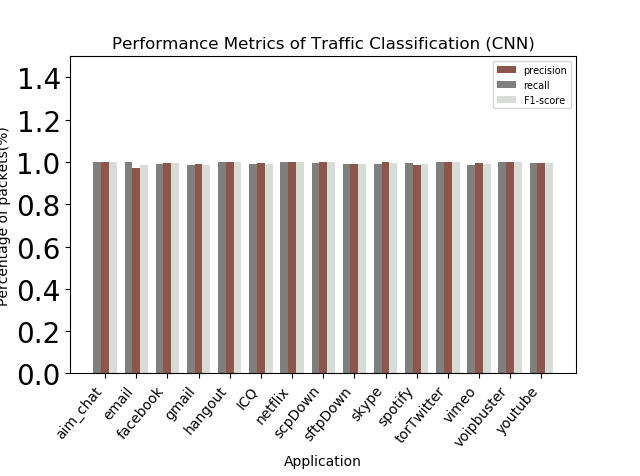}
		\end{minipage}%
	}%
	
	\caption{Performance metrics of DL-based encrypted traffic identification method on ISCX}
	
	\centering
	\subfigure[ROS]{
		\begin{minipage}[t]{0.25\linewidth}
			\centering
			\includegraphics[width=1.7in]{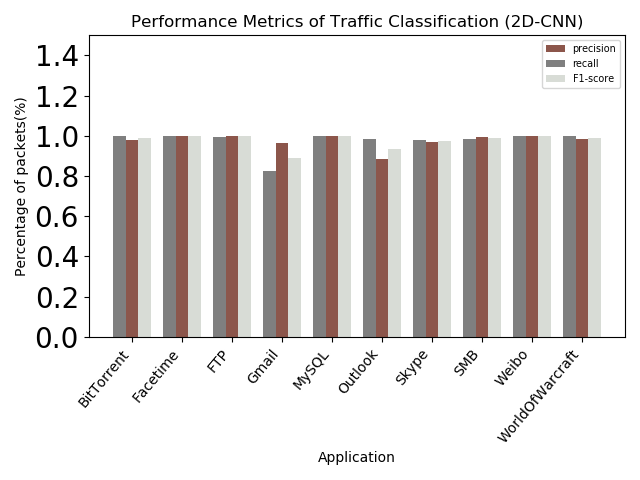}
		\end{minipage}%
	}%
	\subfigure[SMOTE]{
		\begin{minipage}[t]{0.25\linewidth}
			\centering
			\includegraphics[width=1.7in]{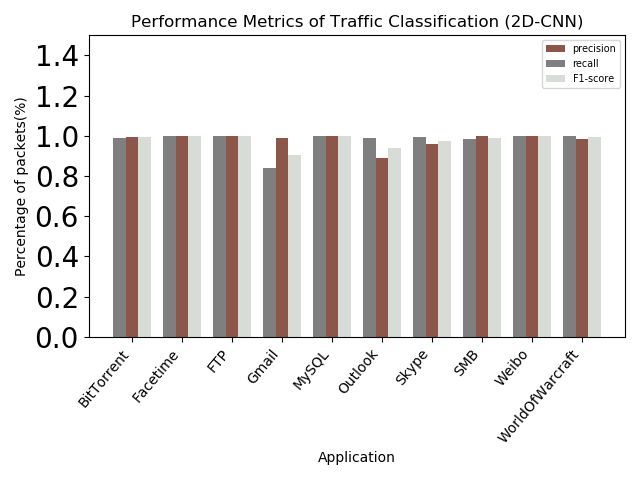}
		\end{minipage}%
	}%
	\subfigure[GAN]{
		\begin{minipage}[t]{0.25\linewidth}
			\centering
			\includegraphics[width=1.7in]{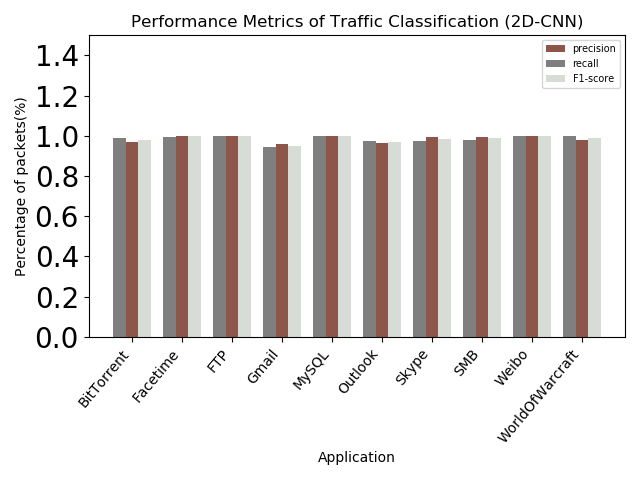}
		\end{minipage}%
	}%
	\subfigure[PacketCGAN]{
		\begin{minipage}[t]{0.25\linewidth}
			\centering
			\includegraphics[width=1.7in]{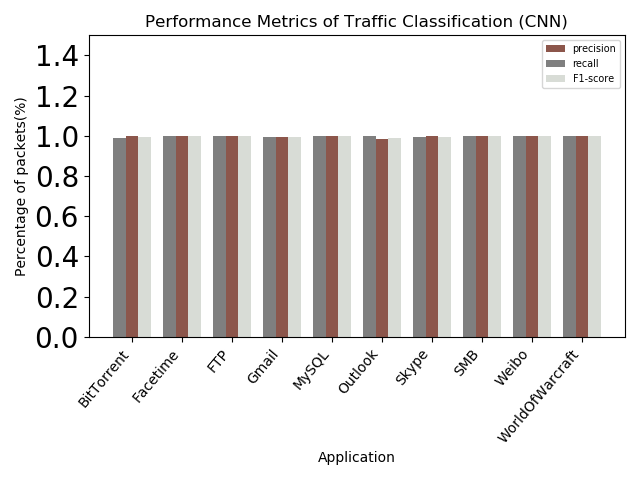}
		\end{minipage}%
	}%
	\caption{Performance metrics of DL-based encrypted traffic identification method on USTC-TFC2016}	
	\label{performance}
\end{figure*}

Fig~\ref{performance} shows the performance metrics of 8 datasets more clearly. As we can see from Fig~\ref{performance} that PacketCGAN greatly outperforms the other three methods for the minor class like gmail, facebook.

From Table ~\ref{tab:Desc_Samples}, we can see the detail of the evaluation performance metrics of classification over the datasets augmented by different methods more clearly. For simplicity, only the performance metrics of ISCX2012 datasets are presented here as an example. The performance statistics of classification based on all four methods are shown in Table ~\ref{table:performance_classifier}, apparently, our proposed PacketCGAN has achieved better performance than the other three methods.

\begin{table*}[htb]	
	\centering  
	\fontsize{5.5}{8}\selectfont  
	\begin{threeparttable}  
		\caption{Description of Experimental results on different generative methods for ISCX2012}  \label{tab:Desc_Samples}  
		\begin{tabular}{l|ccc|ccc|ccc|ccc|ccc|}  
			\toprule  
			\multirow{2}{*}{\textbf{Application}}&
			\multicolumn{3}{c}{\textbf{Unbalanced dataset}}&\multicolumn{3}{|c}{\textbf{ROS  dataset}}&\multicolumn{3}{|c}{\textbf{PacketCGAN  dataset}} &\multicolumn{3}{|c}{\textbf{GAN  dataset}}
			&\multicolumn{3}{|c}{\textbf{PacketCGAN  dataset}}
			\cr  
			\cmidrule(lr){2-4} \cmidrule(lr){5-7}  \cmidrule(lr){8-10}  \cmidrule(lr){11-13}  \cmidrule(lr){14-16}
			&\textbf{Precision} & \textbf{Recall} & \textbf{F1-Score} &\textbf{Precision} & \textbf{Recall} & \textbf{F1-Score}&\textbf{Precision} & \textbf{Recall} & \textbf{F1-Score}&\textbf{Precision} & \textbf{Recall} & \textbf{F1-Score}&\textbf{Precision} & \textbf{Recall} & \textbf{F1-Score}\cr  
			
			\midrule  
			AIM			 & 0.8823&0.9027&0.8924  	& 0.9757&0.9557&0.9656 &	0.9532&0.9386&0.9642  & 0.9989&0.9989&0.9989 &	0.9792&0.9875&0.9833\cr
			Email-Client & 0.9949&0.9893&0.9921 & 0.9985&0.9866&0.9925 &	0.9541&0.9627&0.9558  & 0.9821&0.9584&0.9701 &	0.9792&0.9875&0.9833\cr
			Facebook	 & 0.9287&0.913&0.9208 & 0.9808&0.9571&0.9688 &	0.9531&0.9639&0.9579   	   & 0.9469&0.9667&0.9567 &	0.9792&0.9875&0.9833\cr
			Gmail		 & 0.9803&0.988&0.9841 & 0.9939&0.9919&0.9929 &	0.9426&0.9578&0.9512 	      & 0.9417&0.9436&0.9427 &	0.9792&0.9875&0.9833\cr
			Hangout		 & 0.9421&0.9827&0.962 & 0.9712&0.9881&0.9796 &	0.9578&0.9621&0.9601       & 0.9669&0.9881&0.9774 &	0.9792&0.9875&0.9833\cr
			ICQ			 & 0.9778&0.9225&0.9494 & 0.9343&0.9807&0.9569 &	0.9731&0.9639&0.9681     & 0.9731&0.9778&0.9754 &	0.9792&0.9875&0.9833\cr
			Netflix		 & 0.9999&0.9965&0.9982 & 0.9983&0.9978&0.998 &	0.9781&0.9651&0.9688         & 0.998&0.9998&0.9998 &	0.9792&0.9875&0.9833\cr
			SCP			 & 0.9997&0.9989&0.9993 & 0.9343&0.9982&0.9991 &	1&0.9679&0.9736  	       & 0.9891&0.9872&0.9882 &	0.9792&0.9875&0.9833\cr
			SFTP		 & 0.9963&0.9942&0.9952 & 0.9995&0.998&0.9988 &	0.9841&0.9837&0.9826  	   & 0.9833&0.9652&0.9742 &	0.9792&0.9875&0.9833\cr
			Skype		 & 0.9473&0.992&0.9692 & 0.996&0.996&0.996 &	0.9881&0.9871&0.9873  	      & 0.9787&0.9787&0.9787 &	0.9792&0.9875&0.9833\cr
			Spotify		 & 0.995&0.9935&0.9942 & 0.9952&0.9975&0.9964 &	0.9731&0.9652&0.9663  	   & 0.9768&0.9671&0.9719 &	0.9792&0.9875&0.9833\cr
			torTwitter & 0.9343&0.9993&0.9997 & 0.9343&0.9992&0.9996 &	0.9831&0.9846&0.9841  	 & 0.9908&0.9998&0.9954 &	0.9792&0.9875&0.9833\cr
			Vimeo		 & 0.9973&0.9951&0.9962 & 0.9948&0.9973&0.9961 &	0.9656&0.9631&0.9642  & 0.9718&0.9564&0.9641 &	0.9792&0.9875&0.9833\cr
			voipbuster & 0.9991&0.9965&0.9978 & 0.9343&0.9909&0.9954 &	0.9639&0.9745&0.9721     & 0.9887&0.9826&0.9856 &	0.9792&0.9875&0.9833\cr
			Youtube		 & 0.9975&0.9986&0.998 & 0.9997&0.9982&0.999 &	0.968&0.9662&0.9731  		& 0.9613&0.9807&0.9709 &	0.9792&0.9875&0.9833\cr						
			\midrule
			
			Average&{\bf 0.9759}&{\bf 0.9775}&{\bf 0.9766}&{\bf 0.9892}&{\bf 0.9889}&{\bf 0.9891}&{\bf 0.9751}&{\bf 0.9789}&{\bf 0.9771}&{\bf 0.9766}&{\bf 0.9767}&{\bf 0.9766}&{\bf 0.9936}&{\bf 0.9958}&{\bf 0.9947}
		\end{tabular}  
	\end{threeparttable}  
\end{table*}

\begin{table}[htb]
	\centering
	\fontsize{6.5}{8}
	\caption{Classification Performance Statistics.}\label{table:performance_classifier}
	\begin{tabular}{|l|c|c|c|c|}
		\hline
		data augmenting methods & Accuracy & Precision & Recall & F1-Score \\
		\hline
		\textbf{unbalanced dataset}  & 0.9797 & 0.9759 & 0.9775 &  0.9766 \\
		\hline
		\textbf{ROS dataset}   & 0.9889 & 0.9892 & 0.9889& 0.9891 \\
		\hline
		\textbf{SMOTE dataset}   & 0.9769 & 0.9751 & 0.9789& 0.971 \\
		\hline
		\textbf{GAN dataset}   & 0.9766 & 0.9766 & 0.9767 & 0.9766 \\
		\hline
		\textbf{PacketCGAN dataset}   & 0.9951 & 0.9936 & 0.9958  & 0.9947 \\
		\hline
	\end{tabular}
\end{table}

\section{Conclusion and Future work}\label{conclusion}
In this paper, we proposed a CGAN-based traffic data augmenting method called PacketCGAN to solve the class imbalance problem in the dataset. After the specified sample data is generated by the generator of PacketCGAN, the synthesized data is combined with the original data to build  a new balanced dataset. We use MLP/CNN/SAE models to verify the classification performance over different datasets. The experimental results show that the balanced dataset augmented by FlowCGAN can achieve better performance than the others including ROS, SMOTE and GAN. In the future, we will further study other types of GAN applications in traffic classification and  try to solve the problems of unstable training and mode collapse.

\section*{Acknowledgment}
This work was supported by the National Science Foundation of China (61972211)

\ifCLASSOPTIONcaptionsoff
  \newpage
\fi


\renewcommand\refname{Reference}
\bibliographystyle{IEEEtran}
\bibliography{IEEEfull,Reference}

%

%

\begin{IEEEbiography}[{\includegraphics[height=1.25in,clip,keepaspectratio]{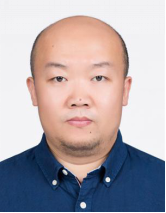}}]{Pan Wang} (M'18) received the BS degree from the Department of Communication Engineering, Nanjing University of Posts\&Telecommunications, Nanjing, China, in 2001, and the PhD degree in Electrical \& Computer Engineering from Nanjing University of Posts\&Telecommunications, Nanjing, China, in 2013. He is currently an Associate Professor in the School of Modern Posts, Nanjing University of Posts\&Telecommunications, Nanjing, China. His research interests include cyber security and communication network security, network measurements, Quality of Service, Deep Packet Inspection, SDN, big data analytics and applications. From 2017 to 2018, he was a visiting scholar of University of Dayton (UD) in the Department of Electrical and Computer Engineering.
\end{IEEEbiography}

\begin{IEEEbiography}[{\includegraphics[height=1.25in,clip,keepaspectratio]{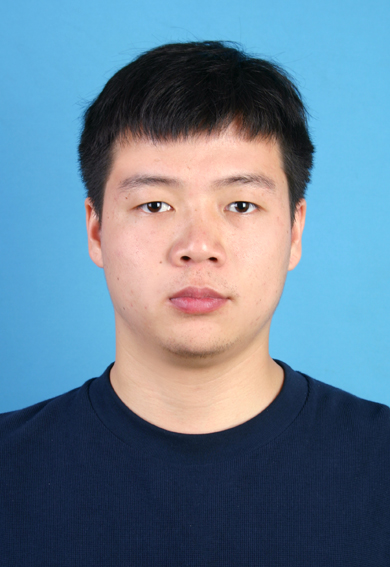}}]{Shuhang Li}  was graduated from Jiangsu University of Science and Technology,Zhenjiang ,China, in 2013. He is currently pursuing a master's degree at Nanjing University of Posts \& Telecommunications, Nanjing China. His research direction is encrypted traffic identification and he also interested in Deep Packet Inspection and applications.
\end{IEEEbiography}

\begin{IEEEbiography}[{\includegraphics[height=1.25in,clip,keepaspectratio]{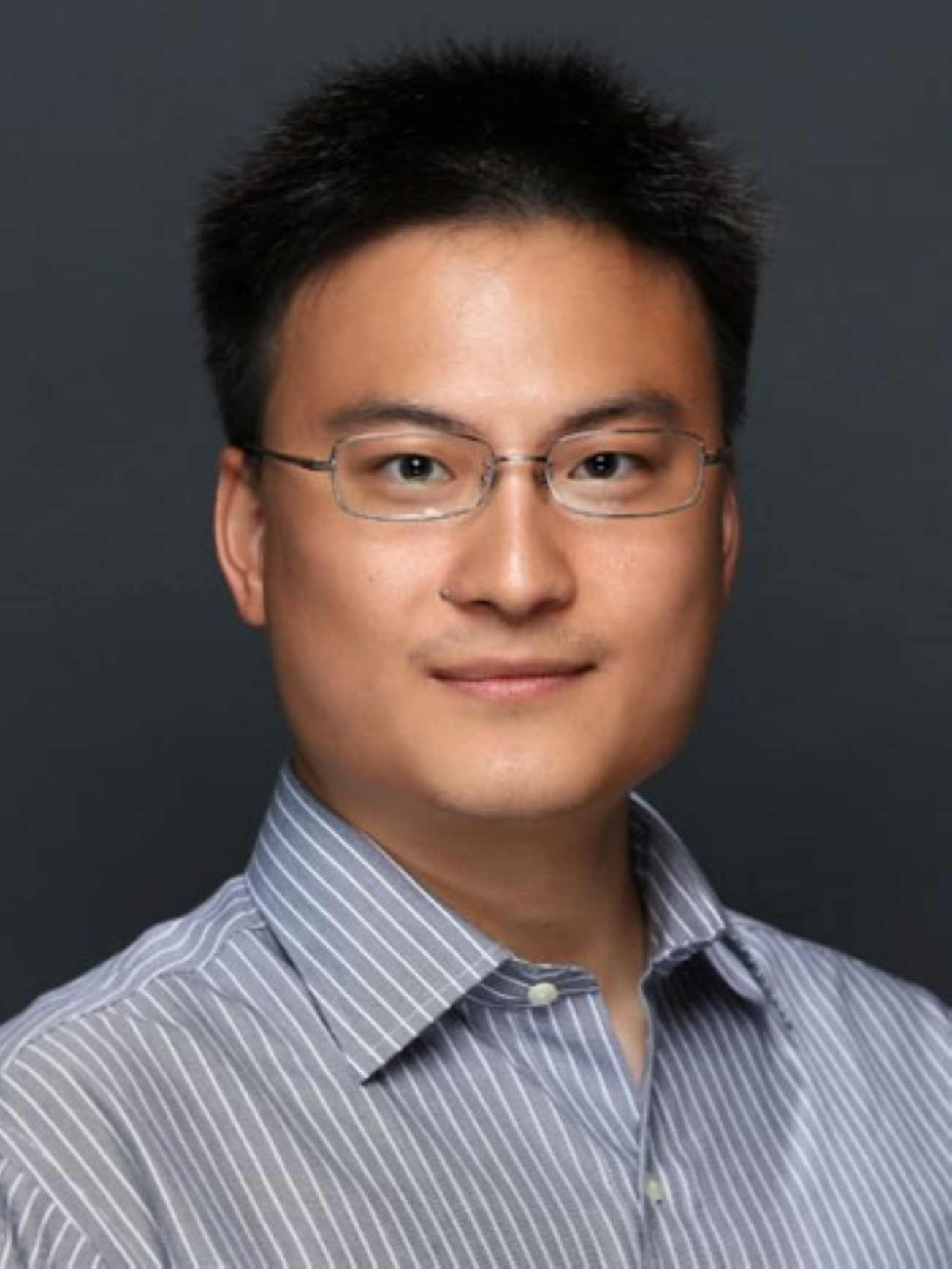}}]{Feng Ye} (S'12-M'15) received the BS degree from the Department of Electronics Engineering, Shanghai Jiao Tong University, Shanghai, China, in 2011, and the PhD degree in Electrical \& Computer Engineering from the University of Nebraska-Lincoln (UNL), NE, USA, in 2015. He is currently an Assistant Professor in the Department of Electrical and Computer Engineering, University of Dayton (UD), Dayton, OH, USA. Prior to joining UD, he was with the Department of ECE, UNL as an instructor and a researcher from 2015 to 2016. His research interests include cyber security and communication network security, wireless communications and networks, green ICT, smart grid communications and energy optimization, big data analytics and applications. He serves as the secretary of the IEEE Technical Committee on Green Communications and Computing (TCGCC). He is currently an Associate Editor of Security and Privacy (Wiley), and China Communications. He serves as the Co-Chair of ICNC'19 Signal Processing for Communications Symposium; the Publicity Co-Chair of IEEE CBDCom 2018; the Co-Chair of Cognitive Radio and Networking Symposium, IEEE ICC 2018. He also serves as a TPC member for numerous international conferences, including INFOCOM, GLOBECOM, VTC, ICC, etc. He is also a reviewer for several IEEE journals, including IEEE Transactions on Big Data, IEEE Transactions on Green Communications and Networking, IEEE Transactions on Smart Grid, IEEE Transactions on Vehicular Technology, IEEE Transactions on Wireless Communications, etc. He is the recipient of the 2015 Top Reviewer of the IEEE Vehicular Technology Society.	
\end{IEEEbiography}

\begin{IEEEbiography}[{\includegraphics[height=1.25in,clip,keepaspectratio]{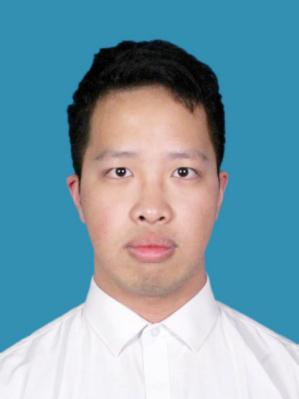}}]{Zixuan Wang}  was born in Nanjing, Jiangsu, China ,in 1994 . He obtained a bachelor's degree from Tongda College of Nanjing University of Posts and Telecommunications in 2017, He is currently pursuing a master's degree in logistics engineering at Nanjing University of Posts and Telecommunications. His research interests include encrypted traffic identification and data balancing.
\end{IEEEbiography}

\begin{IEEEbiography}[{\includegraphics[height=1.25in,clip,keepaspectratio]{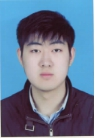}}]{Moxuan Zhang} was graduated from JinLing College of Technology in Nanjing, Jiangsu, China. He is now pursuing a master's degree of computer engineering and data science at Sydney University in Australia. His research interests include encrypted traffic identification and big data.
\end{IEEEbiography}




\end{document}